\documentclass{elsart}
\input{psfig}

\usepackage{amssymb}

\def\astrobj#1{#1}
\def\url#1{{\ttfamily\def\/{/\discretionary{}{}{}}#1}}

\begin{document}

\begin{frontmatter}
\title{Could Edge-Lit Type Ia Supernovae be Standard Candles?}
\author{Enik\H o Reg\H os\thanksref{email}}
\author{Christopher A. Tout}
\address{Institute of Astronomy, The Observatories, Madingley Road, Cambridge CB3 0HA}
\author{Dayal Wickramasinghe}
\address{Astrophysical Theory Centre, Australian National University,
Canberra, ACT 0200, Australia}
\author{Jarrod R. Hurley}
\address{American Museum of Natural History, Central Park West at 79th
Street, New York, U.S.A.}
\author{Onno R. Pols\thanksref{newaddress}}
\address{Department of Mathematics, Monash University, Clayton, Victoria
3168, Australia}
\thanks[email]{E-mail: eniko@ast.cam.ac.uk}
\thanks[newaddress]{Present Address: Astronomical Institute Utrecht,
Utrecht University, P.O. Box 80000, 3508 TA Utrecht, Netherlands}


\begin{abstract}
The progenitors of Type Ia supernovae (SNe~Ia) have not been
identified.  Though they are no longer fashionable we investigate the
consequences if a significant number of SNe~Ia were edge-lit
detonations (ELDs) of carbon/oxygen white dwarfs that have accreted a
critical mass of helium. Our best understanding of the Phillips
relation between light curve speed and peak luminosity assigns both
these phenomena to the amount of $^{56}$Ni produced.  In ELDs there
are two sites of $^{56}$Ni synthesis.  If the peak luminosity is
determined primarily by the CO ratio in the core it is primarily a
function of the initial main-sequence mass of the progenitor of the CO
white dwarf.  If the light curve decay speed is determined by the
total mass of iron group elements ejected this is a function of the
total mass of the ELD at the time of explosion because both the CO
core and the He envelope are substantially converted to $^{56}$Ni.  In
general, binary star evolution ensures that these two masses are
correlated and an empirical relation between peak luminosity and light
curve shape can be expected.  However when we perform population
synthesis for progenitors of different metallicities we find a
systematic shift in this relation that would make distant ELD SNe~Ia
fainter than those nearby.  The abundances of alpha-rich isotopes,
such as $^{44}$Ca, in the solar system indicate that only about 40~per
cent of SNe~Ia could be edge-lit so any systematic effect that could
be present would be correspondingly diluted.  A systematic effect is
still expected even if we examine only the
small subset of ELDs that accrete from a naked helium star, rather
than a He white dwarf.
\end{abstract}

\end{frontmatter}

\section{Introduction}
\label{s:intro}

Luminous type~Ia supernovae (hereinafter SNe~Ia) are amongst the
brightest objects in the Universe.  Observations indicate that their
absolute magnitudes lie in a narrow range, $M_{\rm B} = -18.5\pm 0.5$,
or $\pm 60\,$per cent in luminosity (from Table~1 of Rowan-Robinson
2002).  Subluminous objects such as \astrobj{SN1991bg} (Filippenko et al. 1992)
lie below this range but have quite different spectral features.
Furthermore observations reveal a correlation between maximum absolute
brightness and light curve shape that facilitates an effective
reduction in the standard deviation of absolute luminosities of all
SNe~Ia to $\pm 15\,$per cent (Phillips 1993, Phillips et al. 1999).
This small spread,
coupled with the fact that they can be seen to great distances makes
SNe~Ia excellent standard candles for study of the cosmology of the
Universe.  Observations have determined the rate of expansion with
unprecedented precision (Branch 1998) and have further determined that
this rate is accelerating with time, a measurement that has led to the
invocation of a cosmological constant contributing about $70\,$per
cent of the critical density of the Universe to be added to matter's
contribution of $30\,$per cent (Perlmutter et al. 1999, Riess et
al. 1998).  Further credence is lent to this result by
cold-dark-matter models of the angular variation of the cosmic
microwave background (de Bernardis et al. 2000) and gravitational
lensing measurements (Wittman et al. 2000). Taken together, these
results strongly favour a universe dominated by a new form of energy,
as yet unexplained.

Each of the steps leading to the above conclusions needs to be
critically analysed.  In this paper we look at the assertion that
SNe~Ia can be used as standard candles in the light of what we know
from current state-of-the-art binary population synthesis
calculations.  These calculations point to a range of possibilities
for the progenitors of SNe~Ia, none of which can be eliminated with
any certainty.  Likely candidates are double-degenerate
super-Chandrasekhar white dwarf mergers, single-star
Chandrasekhar-mass CO white dwarf ignitors, and edge-lit detonations
(ELDs) of carbon/oxygen white dwarfs that have accreted a critical
mass of helium.  The relative populations predicted for these types
are model dependent, although our calculations (Hurley, Tout \& Pols
2001) predict a dominant population of ELDs.  Though only a few
attempts have been made to calculate spectra of edge-lit detonations
(H\"oflich and Khokhlov 1996, Nugent et al. 1997) these find that the
synthetic light curves do not match the observations as well as those
for the standard model of carbon ignitors and that the spectra are too
blue.  However, the possibility that the ELDs may be a significant
contributor to SNe~Ia cannot be ruled out on these grounds at present
because a comprehensive set of explosion calculations are presently
unavailable (Branch 2000) and alternatives may indeed be found to fit
the data.

In this paper we point out that, if a significant fraction of SNe~Ia
were ELDs, there is a systematic effect that can be identified which
could make the more distant SNe~Ia fainter than the nearby.  The
evidence for an accelerating universe from SNe~Ia, which requires they
be standard candles, would then be undermined.  The abundance of alpha
rich isotopes such as $^{44}$Ca in the solar system suggests that perhaps
40~per cent of SNe~Ia could be ELDs.

\section{Binary Star Progenitors}

The major energy source of SNe~Ia is the decay of $^{56}$Ni to
$^{56}$Fe and the total energy released in a SN~Ia is consistent with
the decay of approximately a solar mass of $^{56}$Ni.  These facts
strongly implicate the thermonuclear explosion of a white dwarf
though the actual explosion mechanism is not fully understood
(Hillebrandt \& Niemeyer 2000).
White dwarfs may be divided into three major types: (i)~helium white
dwarfs, composed almost entirely of helium, form as the degenerate
cores of low-mass red giants ($M\le2\,M_\odot$) which lose their
hydrogen envelope before helium can ignite; (ii)~carbon/oxygen white
dwarfs, composed of about $20\,$per cent carbon and $80\,$per cent
oxygen, form as the cores of asymptotic giant branch stars or naked
helium burning stars that lose their envelopes before carbon ignition
(with progenitors typically in the range $1-6\,M_\odot$); and
(iii)~oxygen/neon white dwarfs, composed of heavier combinations of
elements, form from giants that ignite carbon in their cores but still
lose their envelopes before the degenerate centre collapses to a
neutron star (with progenitors typically in the range $6-8\,M_\odot$).
\par
An overview of the evolution of close binary systems is given by
Hurley, Tout \& Pols (2002; and references therein) and we review here
those aspects which are directly relevant to SNe~Ia.
In such systems, mass transfer can increase the mass of a
white dwarf.  Close to the Chandrasekhar mass ($M_{\rm Ch}\approx
1.44\,M_\odot$) degeneracy pressure can no longer support the star
which collapses releasing its gravitational energy.  The ONe white
dwarfs lose enough energy in neutrinos and collapse sufficiently,
before oxygen ignites, to avoid explosion (accretion induced collapse,
AIC).  The CO white dwarfs, on the other hand, reach temperatures
early enough during collapse for carbon fusion to set off a
thermonuclear runaway under degenerate conditions and release enough
energy to create a SN~Ia.  Accreting He white dwarfs reach
sufficiently high temperatures to ignite helium well below $M_{\rm
Ch}$ ($M\approx 0.7\,M_\odot$, Woosley, Taam \& Weaver 1986) but an
explosion under these conditions is expected to be quite unlike a
SN~Ia.
\par
The process is further complicated by the nature of the accreting
material.  If it is hydrogen-rich, accumulation of a layer of only
$10^{-4}\,M_\odot$ or so leads to ignition of hydrogen burning
sufficiently violent to eject most, if not all of or more than, the
accreted layer in the well known novae outbursts of cataclysmic
variables (Warner 1995).  The white dwarf mass does not significantly
increase and ignition of its interior is avoided.  However if the
accretion rate is high $\dot M > 10^{-7}\,M_\odot\,\rm yr^{-1}$
hydrogen can burn as it is accreted, bypassing novae explosions
(Paczy\'nski \& \.Zytkow 1978), and allowing the white dwarf mass to
grow.  Though, if it is not much larger than this, $\dot M > 3\times
10^{-7}\,M_\odot\,\rm yr^{-1}$, hydrogen cannot burn fast enough so
that accreted material builds up a giant-like envelope around the core
and burning shell which rapidly leads to more drastic interaction with
the companion and the end of the mass transfer episode.  Rates in the
narrow range for steady burning are found only when the companion is
in the short-lived phase of thermal-timescale expansion as it evolves
from the end of the main sequence to the base of the giant branch.
Super-soft X-ray sources (Kahabka \& van den Heuvel 1997) are probably
in such a state but, without invoking a special feedback mechanism
(Hachisu, Kato \& Nomoto 1996), cannot be expected to remain in it for
very long and white dwarf masses very rarely increase sufficiently to
explode as SNe~Ia.  This however remains the currently favoured
progenitor, though we note that the continuing absence of a
detection of hydrogen that may have been stripped from the mass-losing
companion, either at early stages (Cumming et al. 1996) or late (Eck et 
al. 1995), may change this popular opinion in the future.
\par
At first sight, a more promising scenario might be mass transfer from
one white dwarf to another.  In a very close binary orbit
gravitational radiation can drive two white dwarfs together until the
less massive fills its Roche lobe.  If both white dwarfs are CO and
their combined mass exceeds $M_{\rm Ch}$ enough mass could be
transferred to set off a SN~Ia.  However if the mass ratio $M_{\rm
donor}/M_{\rm accretor}$ exceeds $0.628$ mass transfer is dynamically
unstable because a white dwarf expands as it loses mass.  Based on
calculations at somewhat lower, steady accretion rates, Nomoto \& Iben
(1985) have claimed that the ensuing rapid accretion of material
allows carbon to burn in mild shell flashes, converts the white dwarf
to ONe and ultimately leads to AIC and not a SN~Ia.  For smaller
mass ratios accretion proceeds on the gravitational radiation
timescale and in some cases $M_{\rm Ch}$ can be reached.
\par
If a CO white dwarf accretes from a He white dwarf the mass ratio is
generally small enough for dynamically stable mass transfer so that a
helium layer builds up on the surface of the CO white dwarf.  As in
the nova explosions of hydrogen, the base of this helium layer
eventually reaches a temperature at which helium can ignite in a
degenerate flash.  Unlike the novae this requires about
$0.15\,M_\odot$ of helium (Woosley \& Weaver 1994).  It can be
envisaged that ignition of such a massive helium layer can detonate
the CO core either by compressing it or by an inwardly propagating
heating front (Branch \& Nomoto 1986; Livne \& Glasner 1990).  Typical
total masses of these edge-lit detonations (hereinafter ELDs), though
below $M_{\rm Ch}$, are still of the order of a solar mass so enough
energy exists to explode the star and enough $^{56}$Ni can be formed
to make a SN~Ia.
\par
Our population synthesis predicts, in general, more
ELDs than merging CO white dwarfs by typically a factor of six
(twenty-eight if only dynamically stable mass transfer is counted).
Note that Kawai, Saio \& Nomoto (1987) investigated helium accreting
spherically on to a CO core and claimed that helium can burn
non-degenerately if accreted steadily at rates above about $3\times
10^{-8}\,M_\odot\,\rm yr^{-1}$ and so avoid setting off an ELD.
Because gravitational radiation drives mass transfer from a helium
white dwarf donor at a rate considerably in excess of this,
progenitors involving two white dwarfs have been ruled out in the
past.  However, if accretion is through a disc the CO core is able to
cool as it accretes and a degenerate layer of helium can still build
up.
\par
Spectral analysis of SNe~Ia has also been inconclusive.  The spectra
of the very luminous \astrobj{SN~Ia 1991T} are consistent with $^{56}$Ni
production both at its centre and in a shell at the outside (Liu,
Jeffery \& Schultz 1997, Fisher et al. 1999).  Though Fisher et
al. (1999) favour a super-Chandrasekhar mass model, this is consistent
with the two sites of thermonuclear runaway, the helium envelope and
the CO core, present in an ELD.  A potential problem with any claim
that the majority of SNe~Ia are ELDs is the lack of helium found in
their spectra (Mazzali \& Lucy 1998).  However there remains
sufficient uncertainty in the explosion models, such as exactly how
much helium survives the explosion (typically less than
$0.08\,M_\odot$ in the calculations of Livne \& Arnett, 1995), and
the conversion to observed spectra, in particular the assumption that
the helium shell remains spherically symmetric around the exploding CO
core, that they cannot be ruled out with certainty.  This problem is
no more severe than is the lack of evidence for hydrogen for the
currently favoured single degenerate models.  Indeed ELDs have
received almost no attention in recent years and it may be worthwhile
to revisit them in the light of improved modelling applied to the
standard model.

\section{Population Synthesis}

\begin{table*}
\caption{Supernovae rates for various progenitors per $1{,}000\,$yr
and their yields.}
\label{tab1}
\begin{tabular}{@{}lccccc}
& $\alpha_{\rm CE} = 1$ & $\alpha_{\rm CE} = 3$ & $M_{\rm Ni}/M_\odot$ &
$M_{\rm Ti}/M_\odot$ & $M_{\rm Ca}/M_{\rm Fe}$\\
&\\
Coalescence CO + CO       &  0.05 &  1.20 &  0.76 &  $2.2\times 10^{-6}$ &  $1.9\times 10^{-4}$\\
CO on to CO accretion     &  0.04 &  0.36 &  0.23 &  $6.6\times 10^{-7}$ &  $1.9\times 10^{-4}$\\
&&\\			                                             	    
CO + He wd ELD            &  2.90 &  8.79 &  5.07 &  $3.6\times 10^{-2}$ &  $5.1\times 10^{-3}$\\
CO + naked He ELD         &  0.36 &  1.12 &  0.71 &  $4.3\times 10^{-3}$ &  $4.6\times 10^{-3}$\\
&&\\			                                             	    
He wd + He wd ignition    &  0.15 &  1.39 &  0.63 &  $1.2\times 10^{-2}$ &  $(1.3\times 10^{-2})$\\
He wd + naked He ignition &  0.00 &  0.28 &  0.13 &  $2.5\times 10^{-3}$ &  $(1.3\times 10^{-2})$\\
&&\\			                                             	    
Total CO $> M_{\rm Ch}$   &  0.09 &  1.57 &  0.99 &  $2.8\times 10^{-6}$ &  $1.9\times 10^{-4}$\\
Total ELD                 &  3.26 &  9.91 &  5.79 &  $4.0\times 10^{-2}$ &  $5.0\times 10^{-3}$\\
Total He wd ignition      &  0.15 &  1.67 &  0.75 &  $1.5\times 10^{-2}$ &  $(1.3\times 10^{-2})$\\
&&\\    	    	                                         
Total                     &  3.50 & 13.15 &  7.52 &  $5.5\times 10^{-2}$ &  $5.3\times 10^{-3}$\\
\end{tabular}
\end{table*}

We might hope to gain some insight into the relative sizes of the
progenitor populations from binary population synthesis calculations.
For example we list results from calculations of this type with the
algorithms described by Hurley, Tout \& Pols (2002) in
Table~\ref{tab1}.  The parameter $\alpha_{\rm CE}$ that measures the
efficiency of the common envelope ejection is one of the most critical
and least constrained.  In order to find enough merging CO white
dwarfs to approach the measured rate of $4\pm 2$ SNe~Ia per
$1{,}000\,$yr per galaxy like our own (Cappellaro et al. 1997)
requires $\alpha_{\rm CE}\ge 3$.  In this case only about $20\,$per
cent of merging white dwarfs avoid dynamically unstable mass transfer
and these may be too rare to account for the SNe~Ia
rate observed locally.  However the almost unlimited freedom to vary
such physical parameters as $\alpha_{\rm CE}$ together with all
initial distributions of binary parameters (binary fraction, mass
ratio, period etc.) convinces us that we cannot trust such absolute
numbers from population synthesis.
\par
Given an initial state for a binary system, its masses $M_1$ and~$M_2$ and its
period\footnote{In practice a binary star may have an initial
eccentricity but, in general, tides circularise the orbit before
significant interaction takes place.  Because angular momentum is
conserved during circularisation it is actually the distribution of
semi-latera recta that is appropriate (Hurley, Tout \& Pols 2002).} $P$, and 
a model for all the physical processes of stellar evolution and binary
interaction, we can determine whether or not a
particular system would end its life as a SN~Ia.  These physical processes
include mass transfer both by Roche-lobe overflow or in a stellar
wind, common-envelope evolution, magnetic and gravitational-radiation
braking and all the associated effects on the evolution of the
individual components that comprise the system.  Unfortunately many of 
them are not sufficiently well understood for a precise
quantification.  Particularly troublesome for SNe~Ia is
common-envelope evolution because it is necessary
to bring white dwarfs close enough to interact.
\par
When a red giant or AGB star grows to fill its
last stable potential surface or Roche lobe
it begins to transfer mass
to its companion.  But, as it loses mass, the convective envelope of
a giant expands.  If it is still the more massive component of the
binary, and mass transfer is conservative, the orbit and Roche lobe
shrink.  Consequently the process of mass transfer leads, on a
dynamical timescale, to
the giant overfilling its Roche lobe yet more. The overflow rate
rapidly rises and the
companion, typically a lower-mass main-sequence star, cannot accrete
the material.  Its own Roche lobe is quickly filled and
a common envelope engulfs the whole system.  The two cores, the
relatively dense companion and the core of the giant are then assumed
to spiral together by some, as yet undetermined, frictional
mechanism.  Some fraction of the orbital energy
released is available to drive off the envelope.
If all of it is ejected while the cores are
still well separated we are left with a closer binary system comprising the
unscathed companion and a white dwarf which may evolve to a cataclysmic
variable.  Alternatively, if some of the
envelope still remains when the companion reaches the denser
depths of the common envelope, they can merge leaving a single,
rapidly rotating giant.  Magnetic braking quickly spins
down these merged giants.
\par
Our ignorance of this process is encapsulated in a constant
$\alpha_{\rm CE}$ which measures the fraction of the released orbital
energy that goes into driving off the envelope.  Very few agree on the 
precise definition of $\alpha_{\rm CE}$ and its numerical value is
uncertain to within a factor of ten.  Because additional energy is
always available from the thermal reservoir provided by the nuclear
burning luminosity it may well exceed unity.  Furthermore
it is almost certainly not constant from one case to another (Reg\H os 
\& Tout 1995).
\par
Any white dwarf accreting matter from its companion must have passed
through at least one phase of common-envelope evolution because the
white dwarf must itself have been the core of a giant that could only
have evolved in a much wider system.  Many double degenerate systems
have passed through two phases.
We divide the exploding white dwarfs into three types, (i)~exploding
Chandrasekhar-mass CO white dwarfs, (ii)~ELDs and (iii)~exploding He
white dwarfs.  Each type separates into groups of low and
high intial-period systems.  The high-period systems experience two
common-envelope phases.  The first, when the primary fills its Roche
lobe as a giant, leaves the binary wide enough for the secondary also to
evolve to a giant before it too fills its Roche lobe.  The low-period
systems experience only the second phase.  Mass transfer from the
primary begins while it is crossing the Hertzsprung gap, or perhaps
while it is still on the main sequence.  Mass transfer then proceeds
only on a thermal timescale.  After this Algol phase of evolution we
are left with a white dwarf companion to a more massive secondary in a 
much wider system.  The secondary then evolves to fill its Roche
lobe only as a giant.  Most of the progenitors fall into one of these
two categories but some follow a considerably more convoluted
evolution.  Although we are fully aware of the limitations of
population synthesis
we believe our model to be better than any used by others because of its
fuller treatment of tidal interaction and more careful modelling of
Hertzsprung gap evolution and stellar-wind mass transfer.
\par
A lack of understanding of common-envelope evolution alone ought to be 
enough to make us very wary of the results of population synthesis.
But, like others before us, we go further and calculate the SNe~Ia
rate for the various progenitors in a typical galaxy like our own.  To 
do this we convolve a grid of models (about $5{,}000{,}000$ are
needed to sufficiently resolve the $M_1-M_2-P$ space) with initial
mass functions for each of the components, an initial period
distribution, a binary fraction, a star formation history and a
galactic model.  Although each of these might be reasonably guessed on 
its own, together they give us enough freedom to achieve almost any
result we want even when we require the model to fit observational
constraints on all types of binary star, Algols, cataclysmic
variables, X-ray binaries, symbiotics {\it etc.}  Thus we stress that
the results of binary population synthesis should not be glibly
trusted.

\section{The Supernova Rate}

The observed rate of SNe~Ia is $4\pm 2$ per $1{,}000\,$yr per galaxy
like our own (Cappellaro et al. 1997).  Table~1 gives
the rate for various possible progenitors for two of our population
syntheses that differ only in the value of $\alpha_{\rm CE}$.  We have 
chosen the primary mass $M_1$ from the mass function of Kroupa, Tout
and Gilmore (1993), the secondary mass $M_2$ so as to give a uniform
distribution of mass ratio $q = M_2/M_1$ and the semi-latera recta $l$ from
a flat distribution in $\log l$ with $3 < l/R_\odot < 3\times 10^6$.
The binary fraction and galactic model are condensed into the statement 
that one binary system with $M_1 \ge 0.8\,M_\odot$ forms per year per
galaxy, as is appropriate for our own (Hurley, Tout \& Pols 2002).
The three major groups of progenitors can be further subdivided each 
into two distinct channels.  The exploding Chandrasekhar-mass CO white 
dwarfs can be split into those that have dynamically unstable
(coalescence) and those
that have dynamically stable (accretion) mass transfer.  The ELDs can
be split into those CO white dwarfs accreting from a He white
dwarf and those accreting from a naked helium-burning star.  The
exploding He white dwarfs, which we recall are not likely SNe Ia, can
again be divided into those accreting from a white dwarf and those
accreting from a naked helium star.
\par
It is immediately apparent that by varying $\alpha_{\rm CE}$ alone we
can fit the observed rate with whichever subset of the progenitors we
please.  Because of the two common-envelope phases involved the only
other systems to be so strongly affected by $\alpha_{\rm CE}$ are the
non-interacting double degenerate systems but their numbers are not
yet sufficient to constrain $\alpha_{\rm CE}$ independently (Maxted \&
Marsh 1999).  The well-studied cataclysmic variables usually only
experience a single common-envelope phase in their evolution and so
their numbers, which are uncertain anyway, depend rather more
weakly on $\alpha_{\rm CE}$.  We reiterate that, even without varying
the initial mass and period distributions, which would affect other
types of binary, we are able to fit the SNe~Ia with whichever subset
of the progenitors we please.  The problem is that there are too few
observables to constrain the free parameters of the model.

\section{A Systematic Effect}

The Phillips relation between peak luminosity and light curve decay
speed can be understood (Mazzali et al. 2001) because
both the peak luminosity and the light curve speed depend on the
amount of $^{56}$Ni created in the thermonuclear runaway.  The more
$^{56}$Ni produced, the
brighter the peak while it is the line opacity of the iron peak
elements that determines the speed of the light curve, the more iron
the slower the light curve.  Pinto \& Eastman (2001) explain the
correlation in terms of more $^{56}$Ni leading to a hotter, brighter
supernova but one with a greater dispersion in the velocity of
expanding $^{56}$Ni layers, which delays the photons because 
they must escape a wider range of Doppler-shifted transitions.  What
is important to our argument is that the key parameter, controlling
both peak luminosity and light curve speed, is the nickel mass.
Thus if all the $^{56}$Ni is produced in
a single explosion we expect the correlation irrespective of the nature
of the progenitor.  However in ELDs $^{56}$Ni is formed at two sites,
the He envelope and then the CO core.  Since the latter is still the
most productive site it is still responsible for the peak but
both sites provide opacity that slows the light curve decay.
\par
H\"oflich, Wheeler \& Thielemann (1998) have pointed out that the
precise composition of the CO core, particularly its carbon content,
controls the peak luminosity.  The central composition depends on the
initial, main-sequence mass of the asymptotic giant branch star that
formed the CO white dwarf as its core (e.g. Umeda et al. 1999).  It is 
established during the core helium burning phase before the star has
expanded in radius on the AGB.  If
this mass were independent of the final ELD mass we could not hope to
calibrate them locally.  However, because more massive stars typically
produce more massive remnants we expect some correlation between the
mass of a CO white dwarf and that of its progenitor and so also
between the mass of an ELD and its composition.  The subtleties of
binary star evolution mean that this is not a tight relation because
the total mass of the CO core depends on when the AGB star interacts
with its companion and so on the binary separation.

\begin{figure}
\psfig{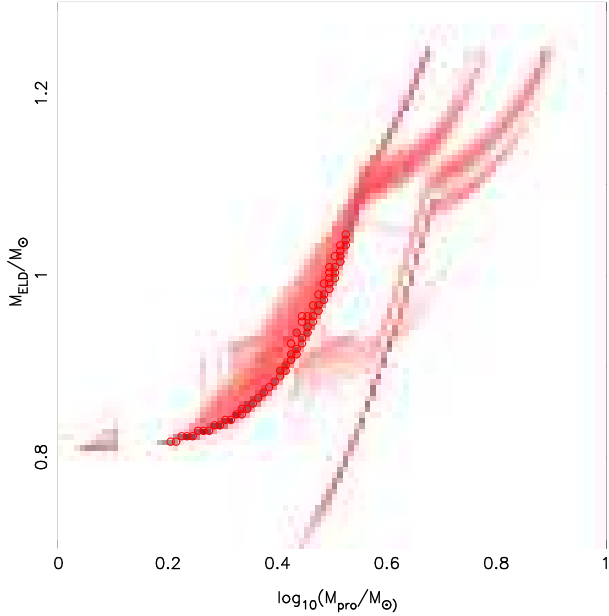}
\caption[]{The correlations between the total mass of ELDs (which
predominantly determines light curve speed) and the mass of the
progenitor of their CO cores (which predominantly determines the peak
luminosity) evolved from stars with metallicity $Z=0.02$ and
$\alpha_{\rm CE} = 3.0$.
The progenitor mass $M_{\rm pro}$ is the initial main-sequence mass of 
the AGB star in which the CO core grew.  It is this that determines the 
C/O ratio.  The
grayscale is determined by the logarithm of the probability of SNe~Ia
falling at a particular point in this plane.  The most common ELDs,
picked out with circles, lie along the lower edge of the top-left
band.  The latter define a narrow (correlated) region so that we can
hope to find correlations between properties of the supernovae for
this metallicity.  The rarer ELDs with naked helium star donors
preferentially occupy the even narrower (more correlated) region to
the lower right.}
\label{fig1}
\end{figure}

\begin{figure}
\psfig{figure=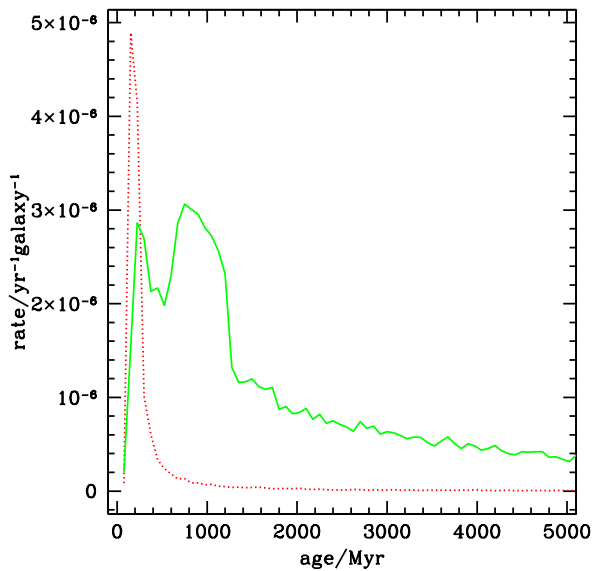,width=80mm,angle=0}
\caption[]{Solid line -- the contribution to the SNe~Ia rate by ELDs
at time $t$ from stars formed at $t-\rm age$.  Dotted line -- the equivalent
contribution from merging pairs of CO white dwarfs.  The ELD rate
peaks less than $10^9\,$yr after the stars formed and the CO merger
rate even sooner.  The narrower ELD peak at this time comes from a
subset of the ELDs with naked helium star donors whose CO components
had massive progenitors ($5-8\,M_\odot$).  It contains none of the
helium white dwarf donors.  These timescales are short compared with the
cosmological age of the Universe at redshifts of $z = 0$ and~$z = 1$,
so that SNe~Ia are essentially a product of the appropriate current
generation of forming stars.  In particular their metallicity is
correlated with their redshift.}
\label{fig2}
\end{figure}

\begin{figure}
\psfig{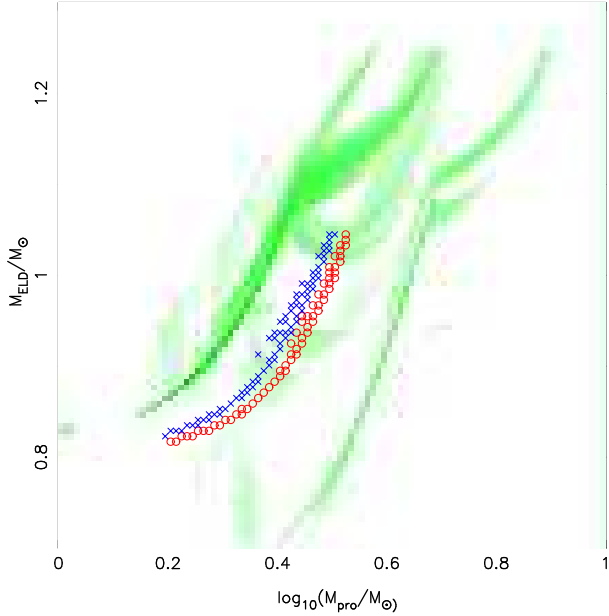}
\caption[]{Grayscale, as in Fig.~\ref{fig1}, of the correlations for
ELDs evolved from progenitors with $Z = 0.001$.  Open circles are the
most common ELDs from progenitors of $Z=0.02$ (as in
Fig.~\ref{fig1}) while crosses are for the similarly most common
ELDs from progenitors of $Z=0.01$.  The systematic offset in these
correlations leads to a systematic shift in the calibration of peak
SN~Ia luminosity by light curve shape.  A given total ELD mass
corresponds to a 30 per cent lower progenitor mass at $Z=0.001$ than
at $Z=0.02$ so that the absolute brightness could be overestimated.}
\label{fig3}
\end{figure}

\par
Fig.~\ref{fig1} is a plot of ELD mass against progenitor mass when $Z
= 0.02$ and $\alpha_{\rm CE} = 3.0$.
There is indeed a correlation with some scatter.  Various less
occupied regions come about because there are many routes by which
stars can reach the final ELD.  Apart from the difference in donor,
the split between long initial period systems with two CE phases and
short-period single-CE systems is most significant.  Though the
details are interesting they are not pertinent to this paper and we
concentrate only on the highly occupied regions.
Let us now suppose
that the progenitor mass predominantly determines the peak luminosity
while the total ELD mass predominately determines the light curve
speed.  The correlation between the two leads to the empirical
relation used to calibrate the peak luminosities, and the scatter in
this relation is due to the variations in formation pathway.
If $\alpha_{\rm CE} = 1$ the region containing the most common ELDs is 
squeezed to the upper left of this distribution and the gap just below 
$M_{\rm pro} = 10^{0.2}$ is filled in but the correlation still exists.
In
practice we expect both the peak luminosity and light curve shape to
depend on some combination of both masses but our arguments are
unchanged.  There is still a one parameter family of SNe~Ia.  Now our
population synthesis reveals that the peak rate of ELDs occurs about
$10^9\,$yr after their progenitors formed (Fig.~\ref{fig2}), just
the time taken for intermediate mass stars to evolve off the main
sequence.  A similar conclusion can be reached by examination of Fig.~2 of
Yungelson \& Livio (2000).
Therefore local SNe~Ia (even those in old galaxies with
currently low star formation rates) have typically population~I
progenitors with a metallicity of $Z = 0.02$.  At a redshift of $z
\approx 1$ the Universe was only one third of its current age and the
local metallicity was lower.  Precisely how much lower is quite
uncertain but our point is most clearly made by comparing a
population~II synthesis with $Z = 0.001$ with that at $Z = 0.02$.
Fig.~\ref{fig3} compares the correlations between ELD and progenitor
masses.  There is a systematic shift in the correlation but the spread
is little changed.  As a result ELDs from population~II progenitors
could appear to have a very similar distribution of peak luminosities
and light curve shapes to population~I but at the same time have a
systematic shift in the relation used to calibrate their peak
luminosities.  Indeed a similar total ELD mass at $Z=0.001$ is now
mapped to a progenitor some 30~per cent less massive than at $Z=0.02$.
Equivalently a certain speed of light curve decay at $Z=0.001$ could
correspond to a significantly fainter peak luminosity than at
$Z=0.02$.  Though $Z=0.001$ is most likely an underestimate of the
metallicity at a redshift of $z = 1$, the effect is still apparent at
$Z=0.01$, which is almost certainly an overestimate, (see
Fig.~\ref{fig3}).
\par
The analysis by Kawai et al. (1987) of helium accreting spherically on 
to a CO white dwarf showed that, at accretion rates above $3\times
10^{-8}\,M_\odot\,{\rm yr}^{-1}$, the helium burns non-degenerately as 
it accretes.  Models we have made with the Eggleton stellar evolution
code (Pols et al. 1995) confirm this and find degenerate ignition to
be possible, even for spherical accretion, if accretion rates are less 
than $10^{-9}\,M_\odot\,{\rm yr}^{-1}$.  Because of their larger size
naked helium stars can feed CO white dwarfs at less than this rate.
In Figs 1 and~3 these ELDs lie preferentially on the narrower
correlation, at higher progenitor mass, to the right of the main
cluster.  This region does not appear to suffer the same systematic shift with
metallicity.  However, even at $Z = 0.01$, the C/O ratio is
significantly higher than at $Z = 0.02$.  For instance models made
with the Eggleton code (Pols et al. 1998) show that a
$2.5\,M_\odot$ progenitor with $Z = 0.01$ produces a CO white dwarf
with central carbon abundance $X_{\rm C12} = 0.15$ compared with
$0.12$ for a star of the same mass but $Z = 0.02$.  A similar abundance
is shared by CO white dwarfs from $2.0\,M_\odot$ progenitors when $Z = 
0.02$.  These conclusions are based
only on our simple arguments that should be tested with detailed
explosion models before any firm deductions can be drawn either way.
They are however sufficient to cast reasonable doubt on the fact that
SNe~Ia are standard candles independent of redshift, at least until we 
can be sure of a sample that is not contaminated with ELDs.

\section{Nucleosynthesis}

We might hope that if we include nucleosynthesis in population
synthesis we can tap into many more observable quantities from stellar
abundances of elements to meteoritic and terrestrial isotopic ratios.
Unfortunately this is not without the expense of introducing new
parameters to our model but we can reduce the ratio of parameters to
observables.  Of particular interest to us for supernovae are the iron
group elements and heavier isotopes that are not synthesized in
earlier phases of stellar evolution such as s-processing in AGB stars.
The greatest uncertainty is introduced by our poor understanding of
the mechanism by which SNe~II explode.  It is not certain how much of
their iron core is trapped in or falls back on to the newly formed
neutron star and how much is expelled to the ISM.
The iron group production by the various core collapse supernovae,
SNIb/c and SNII, is difficult to quantify because it critically
depends on the poorly understood mechanism that drives the explosion.
We can make an estimate based on the work of Woosley \& Weaver
(1995).  Their Table~5 shows a yield that varies from $0.04\,M_\odot$, for a
$12\,M_\odot$ to $0.2\,M_\odot$ for a $22\,M_\odot$ star but for most
masses the yield is below $0.1\,M_\odot$.  Above $25\,M_\odot$
their yields vary greatly with their, imposed, piston velocity.  If
black-hole remnants are common the yields are very small.  In any
case these very massive progenitors ought to be much rarer.  We use
an average yield of about
$0.08\,M_\odot$ of Fe (mostly as $^{56}$Ni) expelled to the ISM
per SN~II as a reasonable but rather uncertain estimate.  Because
$^{56}$Ni decay is the major source of energy in the SNe~Ia their iron
production is much better determined.
Observations (Contardo, Leibundgut \& Vacca 2000) indicate a range in
$^{56}$Ni masses from $0.11\,M_\odot$ for \astrobj{SN1991bg} to $1.14\,M_\odot$
for \astrobj{SN1991T} with an average of $0.58\,M_\odot$ for nine SNe~Ia.  For our 
purposes we would like to know the yield for each of our exploding
white dwarfs and so we appeal theoretical models.
Exploding Chandrasekhar-mass CO
white dwarfs give about $0.63\,M_\odot$ each (Thielemann, Nomoto \&
Yokoi 1986).  The ELD production depends on both CO core mass and He
envelope mass (at least one extra parameter for the model) but the
results of Livne and Arnett (1995) can be fitted quite well by
\begin{equation}
M_{\rm Ni} = 0.75\,M_\odot - 3.0(M_{\rm CO} - 1.0\,M_\odot)^2/M_\odot,
\end{equation}
where $M_{\rm Ni}$ is the total mass of $^{56}$Ni, that will decay to
$^{56}$Fe, produced by an ELD with CO core mass $M_{\rm CO}$.  The mass 
of iron liberated by exploding He white dwarfs would again depend very 
much on the mass and accretion rate.  An appropriate model (Woosley,
Taam \& Weaver 1986) gives $0.45\,M_\odot$ per explosion.  We have
included these yields in our population synthesis calculations and
Table~\ref{tab1} lists the yields per $1{,}000\,$yr per galaxy for the 
model with $\alpha_{\rm CE} = 3$.  If an average SN~II contributes
about $0.08\,M_\odot$ and their rate is about thrice that of SNe~Ia
(Cappellaro et al. 1997) then the iron contribution of SNe~Ia is 
about two and a half times that of SNe~II whatever the true
progenitor.

\subsection{Explosive Helium Burning}

What might distinguish ELDs from exploding Chandrasekhar-mass CO white
dwarfs?  In the latter the thermonuclear runaway is confined to the CO
rich mixture where it produces the $^{56}$Ni and the Si peculiar to
SNe~Ia spectra.  In the ELDs it occurs both in the CO core and in the
He-rich envelope.  In this envelope a small but very significant
fraction of nuclei do not reach the end of the $\alpha$-burning chain
but freezeout as the envelope expands.  Of particular interest are the
heavier isotopes such as $^{44}$Ti and $^{48}$Cr which are not readily
produced elsewhere.  We concentrate on $^{44}$Ti that decays, via
$^{44}$Sc, to $^{44}$Ca in the ISM.  The mass ratio of
$^{44}$Ca/$^{56}$Fe in the Solar System is $1.2\times 10^{-3}$.  An
average SN~II yields at most a ratio of about $6\times 10^{-4}$
(Timmes et al. 1996) and an exploding Chandrasekhar-mass white
dwarf only $3\times 10^{-5}$ (Livne \& Arnett 1995).  As Timmes
et al. point out these two alone cannot account for the Solar-System
abundance and we must turn to explosive He burning.
\par
The model for the exploding He
white dwarf that gave $0.45\,M_\odot$ of iron
(Woosley, Taam and Weaver 1986) yields $8.9\times 
10^{-3}\,M_\odot$ of $^{44}$Ti so that only one such explosion per
eighteen SNe~II or six SNe~Ia would be enough to account for the
Solar-System $^{44}$Ca.  We note however that our calculations with the
Eggleton code show that it is very easy to raise the degeneracy of a
He white dwarf by accretion before the triple-$\alpha$ reaction
begins.  The white dwarf becomes a naked helium burning star and
subsequently a CO white dwarf without any explosive helium burning.
In addition there are no obvious observed candidates for such
explosions, which ought to be almost as bright as SNe~Ia and we might
reasonably discount them altogether.
\par
The contribution from ELDs, not surprisingly, depends, like the Fe
yield, on both the CO core mass and the He envelope mass.  A
fit to the models of Livne and Arnett (1995) is
\begin{equation}
M_{\rm Ti} = \cases{0.0033\,M_\odot + 0.143(M_{\rm CO} -
0.8\,M_\odot)^2/M_\odot & $M_{\rm CO} < 0.8\,M_\odot$\cr
0.0033\,M_\odot & $M_{\rm CO} \ge 0.8\,M_\odot$,}
\end{equation}
where $M_{\rm Ti}$ is the total mass of $^{44}$Ti, that will decay to
$^{44}$Ca produced by an ELD with CO core mass $M_{\rm CO}$.
Again for the model with $\alpha_{\rm CE} = 3$, the mass of $^{44}$Ti
returned to the ISM by each of the various progenitor types is
recorded in Table~\ref{tab1} and the final column gives the
Solar-System ratio if only that progenitor type is combined with
thrice as many SNe~II.  ELDs give a ratio that is four times as large
as is measured
and so are unlikely to be the dominate type of SNe~Ia.  Exploding Chandrasekhar-mass CO
white dwarfs give a ratio over six times too small.  If we exclude the 
CO white dwarfs accreting from He white dwarfs on the grounds that 
they can burn helium non-degenerately then the combination of the ELDs 
accreting from naked helium stars, all the exploding
Chandrasekhar-mass CO white
dwarfs and the SNe~II, with no exploding He white dwarfs,
give a ratio of $2.0\times 10^{-3}$ within a
factor of two of the measured value.  This is therefore our favoured
model based on these calculations.  The progenitors of $40\,$per cent of its
SNe~Ia are ELDs which are unlikely to be standard candles.

\section{Conclusions}

Though they are now quite unfashionable we have investigated whether
or not edge-lit SNe~Ia could behave as standard candles.
We have so identified a systematic effect that could undermine the
cosmological interpretation of SN~Ia measurements at redshift $z
\approx 1$ because it ensures that both nearby and distant SNe Ia
appear to belong to a one parameter family that should obey a
correlation between peak luminosity and light curve shape but at the
same time hides a second parameter, metallicity.  For this
to be important a significant number of SNe~Ia would need to be
helium edge-lit detonations of CO white dwarfs.
Unfortunately we find the results of binary population synthesis (such 
as Hurley et al. 2001) too unreliable to produce absolute numbers.
There are just too many free parameters in the included physics and
the initial distributions of binary properties.  On the other hand the 
relative numbers used in identifying the systematic effects of
metallicity on ELDs do not rely heavily on any of these many
assumptions.  If we include nucleosynthesis in population synthesis we 
find that 40~per cent of SNe~Ia would need to be ELDs if they are the
major source of $^{44}$Ca in the solar system.
\par
Any measurements of the metallicity of supernova progenitors can be
used to test for such effects.  Since we expect the majority of SNe~Ia
to result from relatively recent star formation (our Fig.~\ref{fig2} or
Fig.~2 of Yungelson \& Livio 2000) measurement of the metallicity of
intermediate and massive stars ($M \ge 2\,M_\odot$) in the host galaxy
is also appropriate.  Branch, Romanishin \& Baron (1995) found that
the distribution of peak SNe~Ia luminosities in redder galaxies has a
higher dispersion, before correcting for light curve shape, than that
of those in bluer galaxies.  This is consistent with SNe~Ia in an
older red galaxy coming from a wider range of progenitor metallicities 
but is otherwise inconclusive.  Further investigations of how SNe~Ia
might vary with galaxy type are underway (e.g. Schmidt, private
communication) and we eagerly await the results.
\par
We conclude that we find no evidence that the majority of SNe~Ia
should not be standard candles obeying a one parameter Phillips
relation.  However if any can be identified as ELDs they should be
used neither to establish the Phillips relation nor in the distant
samples.  In particular any object that shows similarities to
\astrobj{SN1991T} should be excluded.

\section*{ACKNOWLEDGMENTS}

We thank Zhanwen Han, Andrew King, Robert Kirshner, Paolo Mazzali,
Philipp Podsiadlowski, James Pringle and Brian Schmidt for stimulating
conversations and helpful comments.
ER and CAT thank the Australian National University and its
Astrophysical Theory Centre for hospitality and generous support.  CAT
is very grateful for an advanced fellowship from PPARC and to
Churchill College for a Lectureship in Mathematics.  DW is grateful to
the UK~PPARC for a visiting fellowship at the IOA, Cambridge.  We also 
thank the various referees who have taken the time to read this work
in detail and provided a wide range of criticisms that have, we
hope, improved our presentation.

\label{lastpage}

\end{document}